\documentclass[a4paper,fleqn,usenatbib]{aa}  

\usepackage{graphicx}	
\usepackage{amsmath}	
\usepackage{xcolor}     
\usepackage{subfig}
\usepackage{supertabular}
\usepackage{caption} 
\usepackage{times}
\usepackage{amsmath,amssymb}
\usepackage{txfonts}
\usepackage{upgreek}

\usepackage[breaklinks=true]{hyperref}
\hypersetup{
	colorlinks = true, 
	urlcolor = blue,  
	linkcolor = blue, 
	citecolor = blue 
}

\newcommand{\Msun}{M$_{\odot}$}
\newcommand{\Lsun}{L$_{\odot}$}

\newcommand{\mum}{$\upmu$m}
\newcommand{\mums}{$\upmu$m \,}
\newcommand{\pp}{$^{\prime \prime}$ \,}
\newcommand{\CIV}{\ion{C}{IV}}
\newcommand{\MgII}{\ion{Mg}{II}}
\newcommand{\CIII}{$\mathrm{C\,III]}$}
\newcommand{\Lya}{Ly${\alpha}$}

\begin{document} 

   \titlerunning{The nature of FIR-bright quasars}
     \authorrunning{E. Hatziminaoglou et al.}
   \title{An ALMA Band 7 survey of SDSS/{\it Herschel}  quasars in Stripe 82:}
   \subtitle{II. The nature of FIR-bright quasars}

   \author{E. Hatziminaoglou
          \inst{1,2,3}\thanks{ehatzimi@eso.org},
          H. Messias\inst{4,5},          
          D. Farrah\inst{6,7}, 
          A. Feltre\inst{8},
          D. Koopmans\inst{9,10}, 
          I. Perez-Fournon\inst{1,2},\\
          R. Souza\inst{11,12},
          L. Wang\inst{6,7} 
          }
   \authorrunning{Hatziminaoglou et al.}
   \institute{Instituto de Astrof\'{i}sica de Canarias, 38205 La Laguna, Tenerife, Spain
   \and
   Departamento de Astrof\'{i}sica, Universidad de La Laguna, 38206 La Laguna, Tenerife, Spain
   \and
   ESO, Karl-Schwarzschild-Str. 2, 85748 Garching bei M{\"u}nchen, Germany
   \and
   Joint ALMA Observatory, Alonso de C\'{o}rdova 3107, Vitacura 763-0355, Santiago, Chile
   \and
   European Southern Observatory, Alonso de C\'{o}rdova 3107, Vitacura, Casilla 19001, Santiago, Chile
   \and
   Institute for Astronomy, University of Hawai'i, 2680 Woodlawn Dr., Honolulu, HI, 96822, USA
   \and
   Department of Physics and Astronomy, University of Hawai'i at Mānoa, 2505 Correa Rd., Honolulu, HI, 96822, USA
   \and
   INAF-Osservatorio Astrofisico di Arcetri, Largo E. Fermi 5, 50125 Firenze, Italy
   \and
   SRON Netherlands Institute for Space Research, Landleven 12, 9747 AD Groningen, The Netherlands
   \and
   Kapteyn Astronomical Institute, University of Groningen, Postbus 800, 9700 AV Groningen, The Netherlands
   \and
   Universidade de Pernambuco, Campus Mata Norte, Nazaré da Mata, PE 55800-000, Brazil
   \and
   Departamento de Física, Universidade Federal Rural de Pernambuco, Recife, PE, 50670-901, Brazil
   }

   \date{Received XXX; accepted XXX}

 
  \abstract
{
We investigate the star formation properties and ultraviolet (UV)/optical spectral characteristics of 142 far-infrared (FIR)-bright quasars at redshifts 1 $\le z \le$ 4 to probe supermassive black hole (SMBH) and  host galaxy co-evolution. By combining multi-wavelength observations  with high-resolution ALMA submillimeter (submm) data, we overcome source confusion and robustly constrain star formation rates (SFRs).
The inclusion of ALMA data to the Spectral Energy Distribution fitting reduces SFR estimates by $\sim$20\% for quasars with single submm counterparts and by a factor $\ge$2 for FIR-blended multi-component systems, while reducing FIR luminosity uncertainties by $\sim$18\%. The revised host SFRs span 500 and 3000 \Msun yr$^{-1}$ (median 1080 \Msun yr$^{-1}$), confirming these systems are extreme starbursts. The very shallow positive correlation between starburst and accretion luminosities suggests that SMBH accretion and star formation in the host are not  coupled on the timescales probed by the observations. 
Compared to the general population of FIR-faint quasars, FIR-bright sources have a factor of three higher detection rates at 1.4 GHz, predominantly following the FIR-radio correlation, and reduced \CIV\ and \Lya\ equivalent widths with unaffected \MgII, suggesting AGN-driven winds that leave \MgII\ relatively unaffected. Together, these results place FIR-bright quasars in a brief transitional phase, bridging heavily obscured ``blowout'' galaxies and optically unobscured blue quasars, where bulge assembly and SMBH growth peak simultaneously during ongoing stellar assembly.}


   \maketitle
%

\section{Introduction}
\label{sec:intro}

The relationship between nuclear activity in galactic centres and star formation in the hosts themselves provides a critical window into the physical mechanisms shaping galaxy evolution. The coupling between stellar growth and supermassive black hole (SMBH) accretion links the build-up of galaxies to that of their central engines,  regulating gas reservoirs and governing the co-evolution of galaxies and SMBHs across cosmic time. Active galactic nuclei (AGN), powered by rapid SMBH accretion, are commonly invoked as agents of suppressed star formation via energetic feedback \citep[see][and more recently \citealt{harrison24}]{fabian12}. In theoretical models, AGN-driven outflows expel the gas required for both star formation and accretion, thereby regulating SMBH growth and establishing observed galaxy–SMBH scaling relations \citep[e.g.][]{dimatteo05, croton06, somerville08, schaye15, weinberger18, dave19, byrne24, martinalvarez25}. This “negative feedback” paradigm is often used to explain the high-mass end of the stellar mass function and the emergence of massive, quiescent galaxies on the red sequence.

However, observational evidence increasingly challenges this simple picture. Intense star formation and SMBH accretion are known to coexist in systems such as local Ultra-Luminous Infrared Galaxies (ULIRGs), as established since the IRAS \citep[e.g.][]{sanders96, farrah03}, ISO  \citep[e.g.][]{genzel98} and {\it Spitzer} \citep[e.g.][]{armus07,hernan11} surveys. Large far-infrared (FIR) surveys with {\it Herschel}/SPIRE \citep[e.g.][]{eales10,oliver12} have extended this view to high redshift, identifying hundreds of quasars out to $z \sim 5$ that exhibit substantial ongoing star formation. In particular, a minority (a few percent) of optically bright SDSS quasars are individually detected in SPIRE bands (hereafter ``FIR-bright'' quasars, in contrast to quasars that are individually undetected in SPIRE bands and  are hereafter referred to as ``FIR-faint'' quasars), and show extreme FIR luminosities ($10^{12}$–$10^{14}$ \Lsun), corresponding to star formation rates (SFRs) often exceeding 1000 \Msun yr$^{-1}$ \citep[e.g.][]{caoorjales12, pitchford16, dong16, kirkpatrick20}. These values exceed, by factors of $\sim$3–10, the average SFRs of the remaining $\sim$95\% of quasars, which are typically FIR-faint and detected only through stacking analysis \citep{mullaney15, ma15, harris16}. Such systems represent some of the most extreme starbursts known. More recently, ALMA and JWST observations have pushed this frontier beyond $z\sim6$, confirming ongoing build up of the host in parallel with accretion onto the central SMBH \citep[e.g.][]{bischetti21,ding23,wilde26}.

The extreme SFRs in FIR-bright quasar hosts place them in the broader context of dusty, infrared-luminous star-forming galaxies, including submillimetre galaxies (SMGs), which dominate the cosmic star formation budget at $1 \lesssim z \lesssim 3$ \citep[e.g.][]{magnelli13,madau14,swinbank14,casey14,dudze20}.
The physical mechanisms driving such intense star formation remain uncertain. Both major mergers and secular cold gas accretion have been proposed, yet neither scenario easily reproduces SFRs exceeding a couple thousand \Msun yr$^{-1}$ without invoking extreme conditions or suppressing feedback entirely \citep{birnboim03, narayanan10, dave10, narayanan15}. Moreover, such SFRs may approach or exceed the “maximum intensity” limit for starbursts, where radiation pressure self-regulates star formation \citep{elmegreen99, geach13}.

Several studies suggest that FIR-bright quasars represent a short-lived evolutionary phase, based on systematic differences in their UV/optical properties compared to FIR-faint quasars. \cite{maddox17} found that {\it Herschel}/SPIRE-detected SDSS quasars without Broad Absorption Lines (BAL) in the H-ATLAS survey \citep{eales10} exhibit larger \CIV\ blueshifts than the broader non-BAL quasar population, consistent with a phase of unobscured, high SMBH accretion and powerful nuclear outflows. 
Similarly, \cite{li20} reported that quasars at $z>5$ detected with SCUBA2 at 850 \mums show weaker \Lya+NV emission than submm-undetected quasars, in line with earlier results for $z>6$ systems \citep[e.g.][]{wang08}. They interpreted this as evidence for an early stage of AGN-galaxy co-evolution, in which the BLR is still developing or partially shielded from the central ionising source, leading to modified emission-line properties. Consistently with this picture and based on the analysis of their X-ray detections in Stripe 82, \cite{kirkpatrick20} identified FIR-bright quasars as a transient evolutionary stage termed ``cold quasars'', a population of optically unobscured, blue quasars hosted by intensely star-forming galaxies. 

Despite the many attempts to reveal the nature of this population, a key limitation in interpreting FIR-bright quasars and their inferred FIR luminosities and SFRs arises from the coarse angular resolution of {\it Herschel}/SPIRE (18\arcsec, 24\arcsec\ and 35\arcsec, at 250 \mum, 350 \mums and 500 \mum, respectively; \citealt{griffin10}), which leads to significant source confusion \citep{wang14, hurley17}. Multiple galaxies may contribute to the observed FIR emission within a single beam, raising uncertainty as to whether the emission originates from the quasar host, physically associated companions, or unrelated foreground/background sources. 

Source confusion also impairs other diagnostics, such as the FIR-radio correlation, the well-established relation between FIR and radio luminosities in star-forming galaxies \citep[e.g.][]{helou85, bell03}. This relation arises because both emissions trace massive star formation: FIR from dust heated by young stars, and radio from synchrotron emission produced by supernova remnants. While the correlation holds across a wide range of galaxy properties, its evolution with redshift remains uncertain. Recent studies suggest mild evolution up to $z \sim 4$ in dusty star-forming galaxies \citep{giulietti22}, whereas radio-quiet quasars, whether lensed \citep[e.g.][]{stacey18,jackson24} or studied via stacking \citep[e.g.][]{retana-montenegro22}, appear to lie on or near the local relation. 

To overcome the limitations imposed by source confusion and the resulting uncertainty in FIR-based SFR estimates, we obtained ALMA Band 7 continuum observations at 0.8\arcsec\ resolution for a statistically significant sample of FIR-bright SDSS quasars in Stripe 82. This effort built on earlier lower-resolution observations with the Atacama Compact Array (ACA), which first highlighted the incidence of source blending in these systems \citep{hatzimi18}. The counterpart identifications presented in \cite[][hereafter Paper I]{hatzimi25} demonstrate that while 65\% of quasars are associated with a single submm source, the remaining 35\% exhibit multiple counterparts within the SPIRE beam, implying that FIR luminosities, and hence SFRs, can frequently be overestimated or misattributed.

In this work, we build directly on these results by exploiting the ALMA-resolved photometry to derive robust, de-confused SFR estimates through updated spectral energy distribution (SED) fitting (Section \ref{sec:sfrs}). This enables a reassessment of the true star-forming properties of FIR-bright quasar hosts and provides a firm foundation for interpreting their nature. We then  
revisit the FIR-to-radio correlation (Section \ref{sec:firc}), which offers an independent constraint on the origin of the infrared emission and the role of star formation. By combining these diagnostics with the findings in Paper I, we move beyond simple luminosity-based classifications and propose a coherent physical picture of FIR-bright quasars (Section \ref{sec:discuss}), assisted by a brief comparison of the UV spectral properties of the objects to those of a matched FIR-faint sample.  
The results presented here provide a critical step toward resolving the long-standing ambiguity surrounding this rare population and place FIR-bright quasars within the broader framework of galaxy evolution.

\section{The ALMA/SDSS quasar sample and available multi-band photometry}
\label{sec:sdsssample}

The sample of 142 ALMA/SDSS quasars, spanning the redshift range from 1 to 4, was drawn from the parent sample of 513 FIR-bright SDSS quasars in the HerMES fields \citep{oliver12}, discussed in detail in \cite{pitchford16}. They were extracted from the quasar catalogues of SDSS Data Releases 7 and 10 \citep[DR7 and DR10, respectively;][respectively]{schneider10,paris14} and lie in in the {\it Herschel} Stripe 82 Survey \citep[HerS;][]{viero14}. For sanity, and as the parent sample was put together more than a decade ago, the 142 quasars were also checked against the SDSS Data Release 16 (DR16) quasar catalogue \citep{lyke20} and the DR16 $ugriz$ photometry is used in the present work. 

For all but one object with available near-infrared (NIR) photometry (136 out of the 142), $J, H$ and $K_s$ data were taken from 
the VISTA Hemisphere Survey (VHS) Data Release 5 \citep{mcmahon21}. As no $Y$-band data from this survey were available for our sources, UKIDSS $Y$-band from Data Release 9 \citep{lawrence13} was used instead, when available (108 out of the 142 sources). 
In addition to the above, 93\% of the objects (134 out of the 142) have Wide-Field Infrared Survey Explorer (WISE) data from AllWISE \citep{write19}. In all cases, missing photometry corresponds to non-detections rather than lack of survey coverage.

To identify the FIR counterparts of the ALMA/SDSS quasars, a 5\pp radius was used to match the optical coordinates with the {\it Herschel}/SPIRE point source catalogues. 
The 350 and 500 \mums fluxes were extracted on the 250 \mums positions, with no cuts applied in signal-to-noise ratio (SNR) for either the 350 or the 500 \mums fluxes. In particular, 61\% (90\%) of the sources are detected at an SNR$\gtrsim 3$ (SNR$\gtrsim 2$) at 350 \mum. These numbers drop to 25\% and 51\%, respectively, for the 500 \mums data. Nevertheless, as these fluxes are prior-based measurements and not individual detections, and since the SED fitting is based on $\chi^2$ minimisation where each photometric data point is weighted by the inverse of its uncertainty, having correspondingly low statistical weight in the fit, all measured fluxes were included in the SEDs, regardless of their SNR. For a detailed description see \cite{pitchford16} and reference therein. HerS was not covered by any of the {\it Herschel}/PACS surveys and no PACS 70, 100 or 160 \mums counterparts were identified in the new {\it Herschel}/PACS Point Source Catalogue \citep{marton24} for any of the sources.

The ALMA in Band 7 (870 \mum) continuum observations were taken under Cycle 9 project 2022.1.00029.S (PI Hatziminaoglou). 
For the sources with more than one submm counterparts within 10\pp from the optical location of the quasar (roughly coinciding with the SPIRE 250 \mums beam), the FIR fluxes were weighted by the relative weights of the 870 \mums fluxes of the submm sources within this radius. The implicit assumption here is that the temperature of the dust in the counterparts are the same as in the primary source and therefore their SEDs have similar shapes. This may or may not be a valid assumption, depending on whether the companions were to be at the redshift of the primary source, but given the lack of spectroscopic data for the secondary counterparts and the proximity of the ALMA Band 7 frequency to the SPIRE bands, we consider this to be a reasonable assumption (for details see Paper I).

The multi-wavelength (0.3 to 870 \mum) SED for 98\% of the 142 quasars is sampled with a minimum of 13 and with up to 17 (for 70\% of the sample) data points. For the remaining three quasars we only have 10 data points for each object. 

Additional ancillary data include 1.4 GHz Very Large Array (VLA) data at an angular resolution of 1.8\pp reaching median rms noise of 52 $\upmu$Jy beam$^{-1}$ in Stripe 82 \citep{hodge11}. This survey is one of the deepest 1.4 GHz surveys conducted over such a large area. Of the 142 ALMA/SDSS quasars, 136 lie in the VLA footprint. To go deeper than the 5$\sigma$ cut of the \cite{hodge11} catalogue, we downloaded cutouts\footnote{\url{https://sundog.stsci.edu/}} of 1\arcmin\ size centred on the optical coordinates of the quasars and measured flux densities at the optical position of the quasars. Forced photometry was performed by fitting a 2D Gaussian with fixed centroid at the position of each source position. Peak flux densities were taken from the fitted amplitude and integrated fluxes were calculated using the Gaussian parameters, pixel scale and beam dimensions, extracted from the FITS headers. Pixel values were converted to mJy using the rms noise reported in the FITS headers. The forced photometry yielded a total of 26 detections at or above 3$\sigma$ among the 136 ALMA/SDSS quasars in the VLA survey's footprint. The fluxes from the forced extraction of the 17 objects present in the \cite{hodge11} catalogue are consistent with the fluxes reported in the catalogue to within $\le$10\%.

\section{Calibration of the FIR-derived star-formation rates}
\label{sec:sfrs}

Approximately 65\% of the quasars in our sample have a single submm counterpart within the SPIRE beam. For these sources, the 870 \mums data point was directly incorporated into their SEDs. For the remaining 35\%, where multiple submm counterparts are present within the SPIRE beam, the SPIRE flux densities were redistributed such that each object was assigned a fraction of the flux in each SPIRE band proportional to the contribution of its associated submm counterpart to the total submm flux within the SPIRE beam. This redistribution of FIR fluxes (hereafter referred to as ``ALMA information'') assumes that, in multiple systems, the relative contribution of each ALMA counterpart is independent of wavelength across the SPIRE and 870 \mums bands, effectively corresponding to similar FIR SED shapes for the individual components. We note that differences in dust temperature or redshift among the counterpart would modify the relative contribution at the SPIRE wavelengths. However, the redshift of the secondary counterparts are unknown, preventing a more physically motivated flux weighting. We therefore adopt this pragmatic approach and acknowledge that it may lead to additional uncertainties FIR luminosities and related quantities of some of the affected sources.

We then applied the simple SED fitting routine described in \cite{fritz06}, developed originally with the specific goal to study the properties of the circumnuclear dust (``torus'') in quasars. In its most general implementation, the code fits the observed fluxes and associated uncertainties using three sets of components: stellar population synthesis (SSP) models, AGN emission, and starburst (SB) templates. However, since the present sample consists exclusively of unobscured, optically bright quasars whose emission dominates over that of their host galaxies across the observed UV-to-NIR range, we restricted the fitting procedure to the AGN and SB components only.

For the AGN component, we employed a subset of 213 smooth torus models from the full suite presented in \cite{feltre12}, which account for both the primary optical emission and the near-to-mid infrared radiation reprocessed by circumnuclear dust.  
The models in the library are precomputed for fixed combinations of the torus parameters (namely dependency of the dust density with the distance from the centre and the angle from the equator, the torus half-opening angle, the optical depth at 9.7 \mums and the outer-to-inner torus radius ratio) and no interpolation between the discrete values is performed during the SED fitting procedure. The reduced grid was selected from the full set of $\sim3700$ models presented in \cite{feltre12}, following the prescription in \cite{hatzimi08}. This approach avoid the oversampling of the model parameter space and the exploration of poorly constrained torus parameters, while retaining the range of SED shapes required to reproduce the UV-to-mid-infrared emission of optically selected quasars. The subset of 213 models was the one used in \cite{pitchford16}.

The SB component was modelled using eight  templates, the bright and faint SMG templates from \cite{dacunha15} and
six semi-empirical templates \citep{berta03}, namely: M82, representative of a ``typical'' starburst; Arp 220, corresponding to a heavily extinguished starburst; NGC 1482, NGC 4102, NGC 5253, and NGC 7714, which span a range of Polycyclic Aromatic Hydrocarbon (PAH) feature strengths, FIR bump widths, and peak wavelengths. 

The free parameters in the SED fitting procedure are the normalisation of the AGN and SB templates, while the physical parameters defining each torus model are discrete quantities associated with each best-fitting template. All torus models are normalised to an intrinsic AGN luminosity of 10$^{46}$ erg s$^{-1}$ and the normalisation required to scale the quasar model to the observed photometry is used to derive the accretion luminosity of the quasar, L$_{acc}$, of each best-fitting solution. Similarly, the SB templates are scaled to derive the starburst luminosity, L$_{\rm SB}$. Note that, although different combinations of torus parameters can produce similar SED shapes, L$_{acc}$ is a lot less affected by these degeneracies, as the main variations among models concerns the detailed shape of the reprocessed emission.

Despite its simplicity, this SED fitting code has been extensively validated and shown to successfully reproduce the UV-to-FIR SEDs of AGN over a broad redshift range \citep[e.g.][]{fritz06, hatzimi08, hatzimi09, pozzi12, feltre13, pitchford16, circosta19, damato20}. 

\begin{figure}
\center
\includegraphics[width=0.5\textwidth]{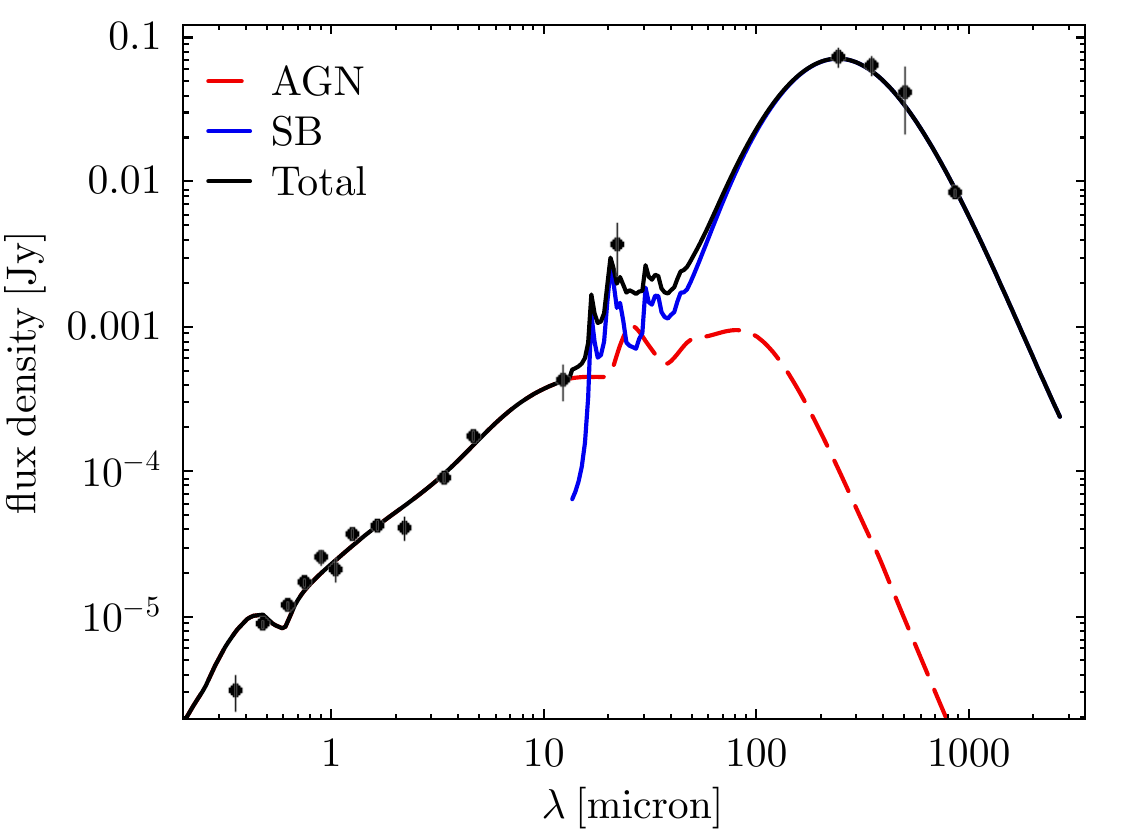}\\
\includegraphics[trim={5.2cm 2.5cm 0cm 2cm}, clip, width=0.635\textwidth]{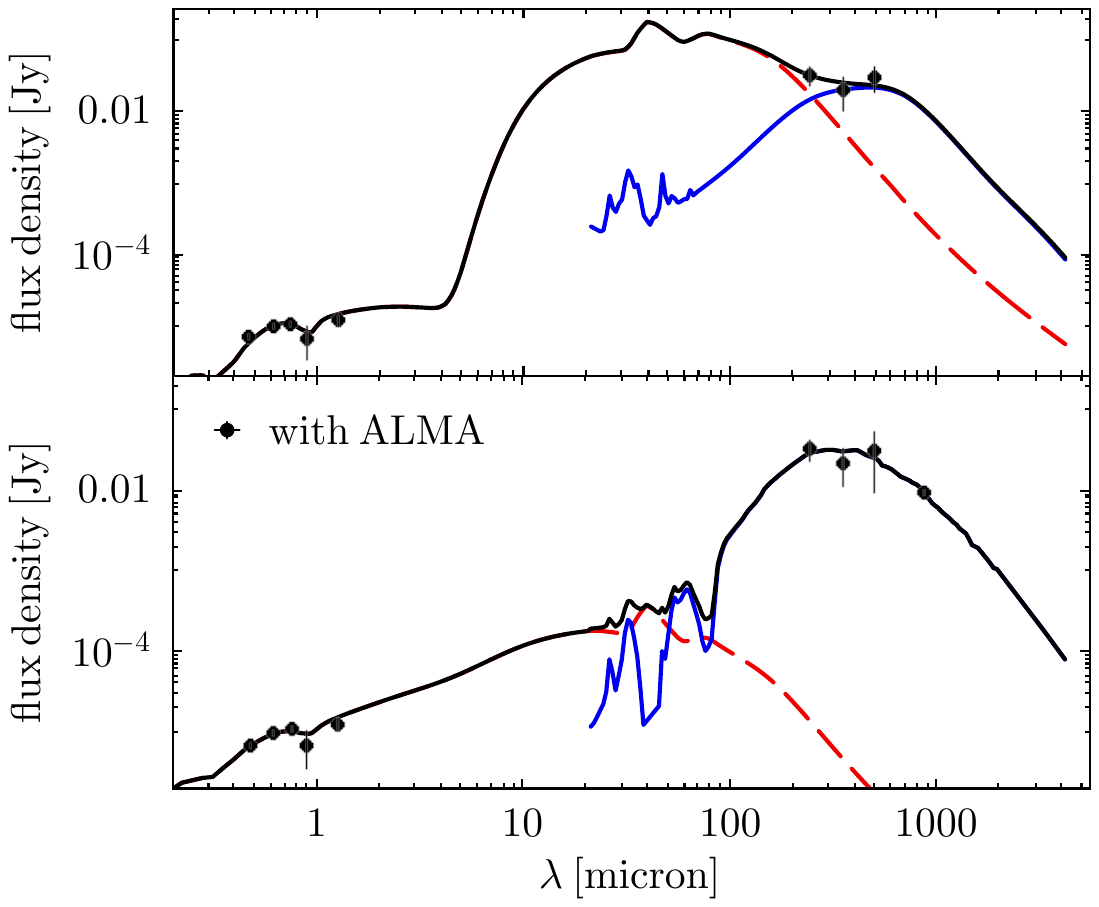}
\caption{SED fitting examples for one object with 17 photometric data points (top) and one with 10 data points, without and with the ALMA information (bottom). The red dashed lines correspond to the AGN component, the blue solid lines to the starburst component. To total SED is shown in black solid line. The wavelengths correspond to the observed values. The data points are shown against observed-frame wavelengths.}
\label{fig:sedfit}
\end{figure}

Two example fits are shown in Fig. \ref{fig:sedfit}, with red dashed, blue and black solid lines corresponding to the AGN, the SB and the total components, respectively. The top panel presents an object with 17 photometric data points, representative of 70\% of the sample. The bottom panel illustrates a rare case with only 10 data points (only three such cases in our sample), where the mid-infrared portion of the SED is entirely unconstrained. While adding the ALMA data point to the SED of the first object had a negligible effect on the outcome of the fitting results, the second example highlights the substantial impact of including the 870 \mums measurement, as discussed below. Both objects have a single submm counterpart within the SPIRE beam. 

For each object, we retain the 213 outputs corresponding to the different AGN models. From these outputs, we construct a $\chi^2$-weighted distribution for each parameter of interest  where the weight of each solution is given by  $w_i$ = exp$(-\chi_i^2/2)$, normalised such that $\sum_i w_i=1$. The best-fit value for a given parameter is taken as the median of the weighted distribution, while the 16th and 84th percentiles define the associated uncertainties.
The values of interest to this work returned by the SED fitting include L$_{acc}$, L$_{\rm SB}$ and the total FIR luminosity, L$_{\rm IR}$ (encompassing the SB and the AGN components), the latter two integrated between 8 and 1000 \mum. The SFR is then computed from L$_{\rm SB}$ following the prescription of \cite{kennicutt98}.

\begin{figure}
\center
\includegraphics[width=0.5\textwidth]{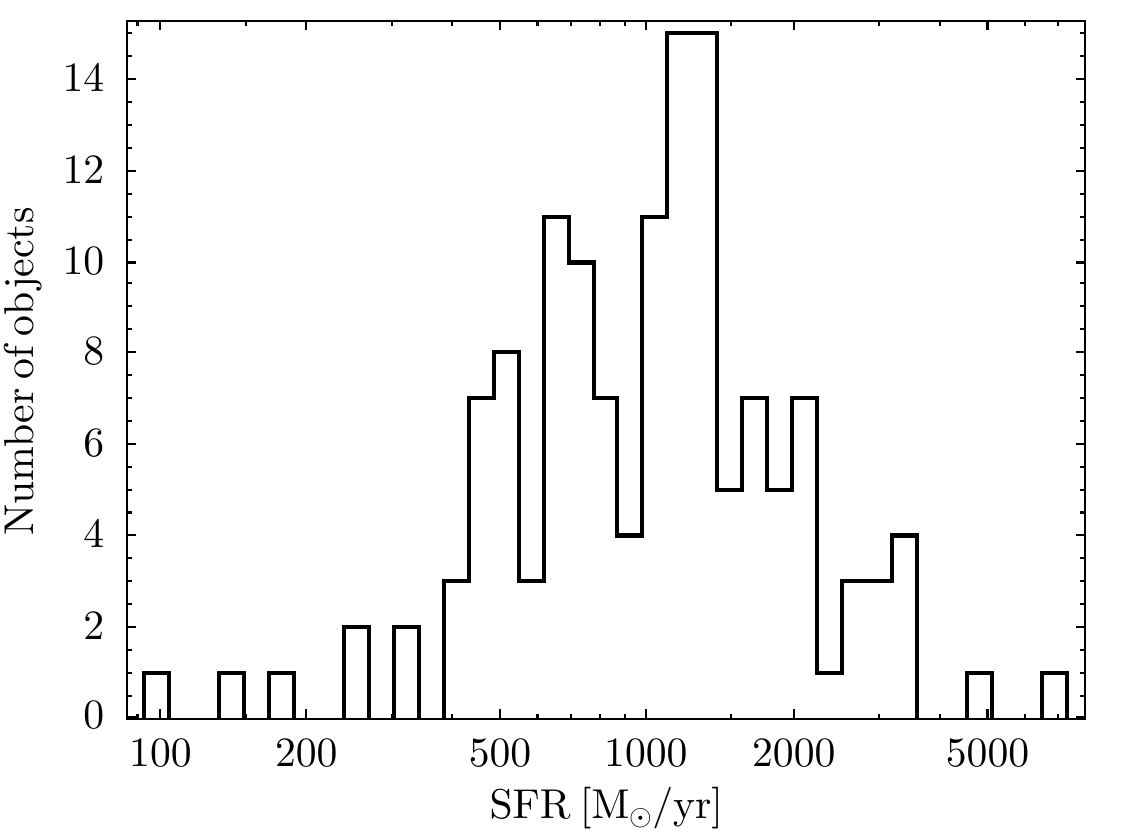}
\caption{SFRs derived from the SED fitting.}
\label{fig:sfrsnoALMA}
\end{figure}

\begin{figure}
    \centering
    \includegraphics[width=0.49\textwidth]{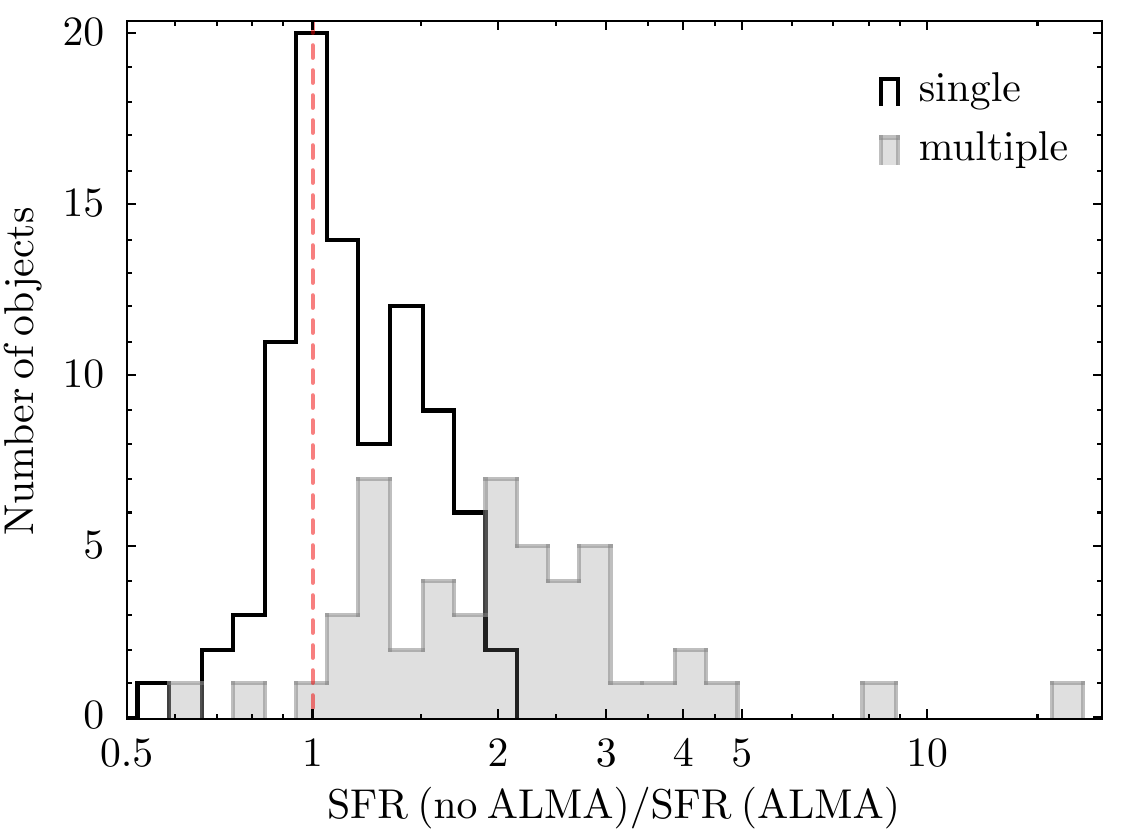}
    \caption{The distribution of the ratio of the SFRs originally derived without ALMA data and those obtained from the SED fitting that includes ALMA measurements (see text for details). Sources with a single ALMA counterpart within the SPIRE beam are shown in black, while those with multiple submm counterparts are shown in grey. Values closer to the red dashed line (ratio of 1) indicate a smaller impact of the ALMA data on the derived SFRs.
    }
    \label{fig:SFRoldnew}
\end{figure}

Figure \ref{fig:sfrsnoALMA} shows the newly derived SFR estimates, after the inclusion of the ALMA information, with a median value of 1080 \Msun yr$^{-1}$, indicative of extreme starburst activity in the hosts of the FIR-bright quasars.
The addition of the 870 \mum \,\ information has a measurable effect on the estimate of the SFRs, as illustrated in Fig. \ref{fig:SFRoldnew}, where the median of the probability distribution function (PDF) for each object is taken as the best value. 
For objects with a single 870 \mums counterpart, the inclusion of the ALMA data point results in systematically lower SFR estimates. Approximately 80\% of the sources exhibit a reduction in the inferred SFRs with a mean decrease of a factor of 1.2.

This is driven by two main factors. First, the ALMA 870 \mums photometry provides much tighter constraints on the long-wavelength end of the SED due to its significantly smaller photometric errors compared to the SPIRE 500 \mums band. Second, the larger SPIRE 500 \mums beam ($\sim$36\arcsec) is more susceptible to source blending than the 250 \mums beam ($\sim$18\arcsec). Despite the lower sensitivity of the 500 \mums channel, this could lead to a systematic overestimation of the 500 \mums flux from the inclusion of neighbouring sources, far beyond the size of the ALMA Band 7 field of view ($\sim$20\arcsec), whereas the high-resolution ALMA data isolates the quasar emission, resulting in more accurate (and lower) SFR estimates. Note that, in some corner cases such as the example shown in the bottom panel of Fig. \ref{fig:sedfit}, the submm data point not only constrains the longer-wavelength portion of the SED, but also enables a more physically plausible fit to the AGN template. As previously mentioned, such cases are rare within our sample, yet they clearly illustrate the importance of the submm data.

For quasars with multiple submm counterparts, adjusting the SPIRE fluxes and adding the 870 \mums point has a stronger effect on the estimated SFRs, lowering the estimates by a factor of $\sim$2.5 on average, but the correction can be a lot more important at times, as seen from the tail of the distribution of the grey histogram in Fig. \ref{fig:SFRoldnew}. 

Perhaps more importantly, including ALMA photometry in the analysis significantly improves the precision of the L$_{\rm IR}$ and the SFRs. We quantify this improvement by comparing the weighted PDF of L$_{\rm IR}$ and SFR for each object in the sample, computed from the SED fitting without ALMA and with ALMA photometry.

For a given parameter $x$ (here L$_{\rm IR}$ or SFR), the weighted mean, $\bar{x}_w$, and weighted standard deviation, $\sigma_w$, are defined as 
\begin{equation*}
\bar{x}_w = \sum_i w_i x_i, \qquad
\sigma_w = \sqrt{\sum_i w_i (x_i - \bar{x}_w)^2}
\end{equation*}
where $x_i$ are the SED fitting solutions and $w_i$ are weights based on the $\chi^2$ values:
\begin{equation*}
w_i = \frac{\exp(-\chi_i^2 / 2)}{\sum_j \exp(-\chi_j^2 / 2)}
\end{equation*}
The fractional reduction of the PDF thanks to ALMA is then
\begin{equation}
\Delta \sigma (\%) = \frac{\sigma_{\rm noALMA} - \sigma_{\rm withALMA}}{\sigma_{\rm noALMA}} \times 100.
\end{equation}
We find that, including ALMA systematically reduces the uncertainties in L$_{\rm IR}$ or SFR. Specifically, the median width reduction is of 17.1\% and 18\% for the  L$_{\rm IR}$ and SFR, respectively, demonstrating that ALMA data help reduce degeneracies in the fitting of the FIR-to-submm part of the SEDs.  

Our SFR estimates are consistent with independent analyses in the literature. Investigating the impact of AGN radio activity on the cold gas reservoirs of quasar hosts of a sample largely overlapping with ours, \cite{wen26} derive SFRs using the Code investigating Galaxy Emission \citep[CIGALE;][]{boquien19} in overall agreement with the values obtained here. Their work, despite being affected by source confusion, provides an independent confirmation of the reliability of our SFR measurements.

\section{Revisiting the FIR-radio correlation}
\label{sec:firc}

A major challenge in assessing the evolution of the FIR-radio correlation is the co-evolution of AGN and dust-obscured star formation during the early phases of galaxy evolution \citep[e.g.][]{delhaize17}, particularly in systems like those examined in this work. 

According to \cite{hodge11}, 7.7\% of SDSS DR7 quasars are detected at 1.4 GHz down to the survey limit (median rms noise of 52 $\upmu$Jy). Among the 99 quasars in our sample originally drawn from SDSS DR7, 20 (38) were detected at 1.4 GHz at or above 5$\sigma$ (3$\sigma$), while the remaining five (nine) $\ge5\sigma$ ($\ge3\sigma$) 1.4 GHz detections are quasars from SDSS DR10. The resulting detection fraction for the FIR-bright quasars in our sample is therefore approximately three times the value reported in \cite{hodge11} for the overall SDSS DR7 quasar population. 

\begin{figure}
    \centering    
    \includegraphics[width=0.49\textwidth]{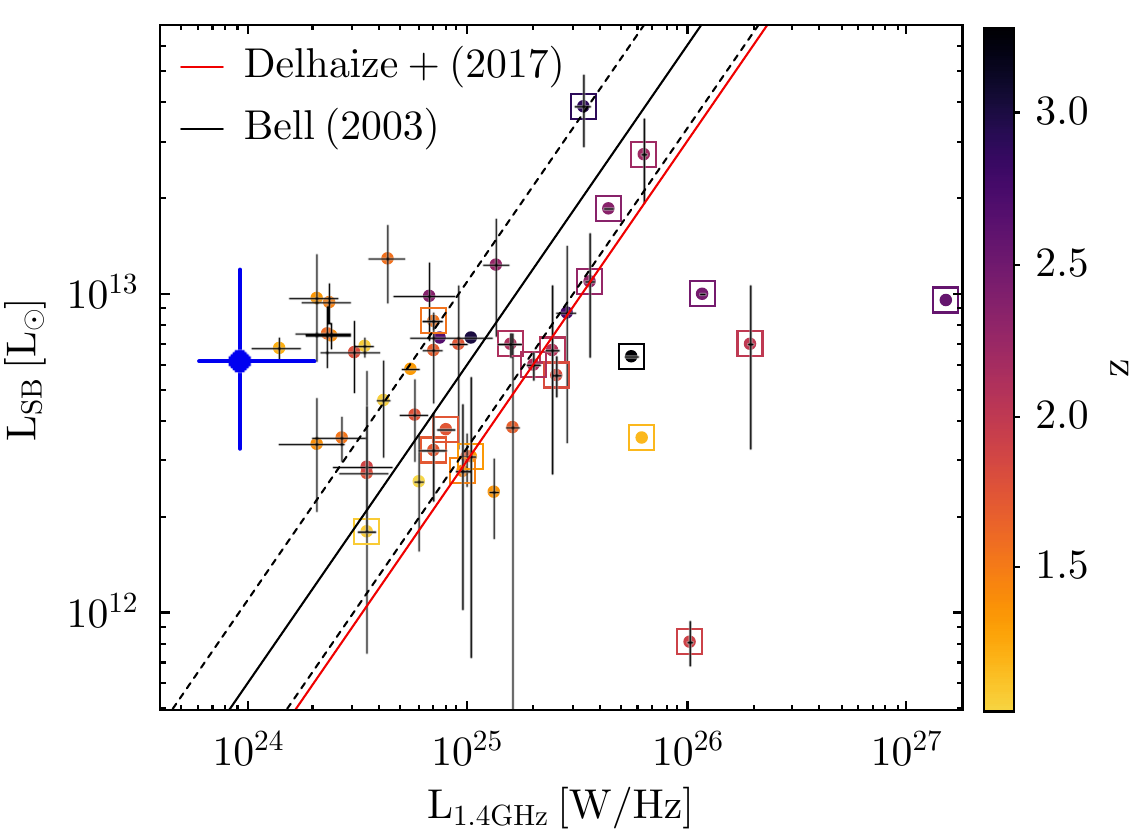}
    \caption{Comparison between L$_{\rm SB}$ and L$_{\rm 1.4GHz}$ for all sources with a forced photometry  detections at 1.4 GHz above 3$\sigma$. Radio detections at or above 5$\sigma$ are indicated by squares. L$_{\rm SB}$ is taken as the median of the SED-fitting posterior distribution, with error bars showing the 16th–84th percentile range. The errors on L$_{\rm 1.4 GHz}$ are the 1$\sigma$ uncertainties derived from the map noise. The objects are colour-coded according to their redshift. The blue point indicates the stacked value for the non-detections. 
    The \cite{bell03} relation for the expected luminosity ratio of ${\rm q_{TIR}=2.64}$ and 0.26\,dex population scatter are shown as a black solid and dotted lines respectively. The relation at $z=2$ from \citep{delhaize17} is shown by the red solid line. 
    }
    \label{fig:lirlrad}
\end{figure}

Figure \ref{fig:lirlrad} shows the starburst luminosity, L$_{\rm SB}$, derived from the SED fitting procedure versus the 1.4 GHz luminosity, L$_{\rm 1.4 GHz}$, obtained adopting a spectral index $\alpha=0.7$, for all objects with a detections at or above 3$\sigma$. The symbols are colour-coded based on the objects' redshift. 
The majority of the 5$\sigma$ radio detections fall within the 1$\sigma$ region of the FIR-radio correlation, indicating that the radio emission originates in the star formation in the quasar hosts. The objects to the right of the \cite{bell03} and \cite{delhaize17} correlations (local and at $z=2$, respectively), are dominated by synchrotron emission from the AGN, and the FIR-radio correlation is not applicable as a star formation tracer. The majority of the 3$\sigma$ detections also fall within or very near the 1$\sigma$ uncertainty of the \cite{bell03} relation.
For the non-detections, we stacked the VLA maps. The blue point in Fig. \ref{fig:lirlrad} indicates the weighted median stack for all non-detections below 3$\sigma$. The stacked median flux is 49.1$\pm$8.9 $\upmu$Jy at a 5.5 $\sigma$ level, just below the VLA survey sensitivity. Stacking all objects below the 5$\sigma$ detection limit yields a stacked flux of 73.8$\pm$7.6 $\upmu$Jy, at a 9.7$\sigma$ detection level.

The high detection fraction of FIR-bright quasars at 1.4 GHz compared to the general SDSS quasar population combined with their proximity to the FIR-radio correlation strongly suggest that their FIR and radio emissions are predominantly powered by the intense star formation in their hosts, rather than by the AGN itself.

\section{Discussion}
\label{sec:discuss}

In this section we discuss the implications of the high SFRs. We then briefly examine the UV/optical properties and compare them with those of a matched sample of FIR-faint quasars. Finally, we bring all the results together to construct a coherent picture, consistent with our findings, that places the rare population of FIR-bright quasars within the broader co-evolutionary scheme of galaxies and AGN.

\subsection{Implications on the FIR-derived SFRs}

Numerous {\it Herschel}-based studies have consistently shown that the AGN do not contribute significantly to the heating of host-galaxy dust in FIR-bright quasars, with the FIR luminosity instead being predominantly powered by star formation \citep[][just to name a few; but see also \citealt{symeonidis17}]{hatzimi10,rosario12,lutz14,stanley15,mullaney15,kirkpatrick15,pitchford16}. More recently, observations with ALMA and JWST have reinforced this picture, while simultaneously revealing the presence of large cold gas reservoirs in the quasar hosts \citep[e.g.][]{wang13,venemans20,mazzucchelli25,silverman26}. These findings provide further support for the extreme SFRs inferred for such systems. 

However, a long-standing uncertainty in {\it Herschel}-based SFRs is the impact of source confusion which can bias FIR measurements, particularly in systems with complex or blended FIR emission. Incorporating high-resolution ALMA photometry into FIR SED fitting has significant implications for deriving SFRs in quasar host galaxies. Although the combined SPIRE+ALMA data primarily constrain the Rayleigh-Jeans of the dust emission rather than its peak, the addition of high-resolution measurements enables the unambiguous identification of the FIR counterparts to the optical quasars. This effectively resolves source confusion within the SPIRE beam, ensuring that the measured FIR emission is correctly associated with the quasar host. As a result, the derived SFRs are systematically lower and more reliable particularly in systems where multiple sources contribute to the blended SPIRE flux (see Section \ref{sec:sfrs}). These findings indicate that previous {\it Herschel}-based SFRs for luminous quasars may have been inflated due to unresolved source blending, in addition to limited wavelength coverage.  Nevertheless, despite these limitations, the revised SFRs confirm that FIR-bright quasar hosts are extreme star-forming systems, with rates comparable to, or exceeding those of, the most intense starbursts known at similar redshifts \citep[e.g.][]{ma15,dacunha15,liao24}.

Beyond establishing the extreme nature of the star formation in the FIR-bright quasars, our analysis also quantifies how the inclusion of high-resolution submm data impacts the robustness of the inferred SFRs. Overall, the $\sim$18\% reduction in the width of the SFR PDFs demonstrates that high-resolution submm data improve the precision of both individual and population-level SFR measurements. This highlights the critical role of spatial resolution even in the absence of direct constraints on the peak of the cold dust emission. A potential caveat, however, arises from the redistribution of SPIRE flux densities, which implicitly assumes a linear proportionality between ALMA and SPIRE emission. This assumption may not always hold, as nearby sources may have different redshifts, dust masses or temperatures. The proportionality would also not hold  in cases with additional physical processes contributing to the observed flux, such as synchrotron emission from the AGN.

Despite the caveats, the above findings reinforce the picture that, FIR-bright quasars are not only sites of rapid black hole accretion but also undergoing intense stellar mass assembly. An additional confirmation comes from the location of these quasars on the L$_{\rm SB}$ - L$_{\rm 1.4 GHz}$ plane indicating that their FIR and radio properties are consistent with those of intensely star-forming galaxies. This also indicates that the physical processes linking star formation to dust heating and cosmic ray production remain consistent even in the extreme environments of radio-quiet quasar hosts. All of the above underscore the coexistence of extreme SMBH growth and vigorous star formation across a wide range of cosmic epochs. 

\subsection{On the co-evolution of quasars and their hosts}
\label{sec:coevolve}

The relation between the recent SFR ($<100$ Myr) and the black hole accretion rate ($\dot{\text{M}}_{\text{SMBH}}$) is a standard probe of the co-evolution of the quasars and their host galaxies. Here we use  L$_{\text{SB}}$ and L$_{acc}$ as proxies for SFR and $\dot{\text{M}}_{\text{SMBH}}$, respectively, given that the two quantities are linearly related through the \cite{kennicutt98} relation and $\dot{\text{M}}_{\text{SMBH}}={\rm{L}}_{acc}/(\eta c^2)$, respectively.
To investigate whether star formation and SMBH growth are directly coupled, a scenario in which more rapidly accreting SMBHs reside in galaxies with systematically enhanced star formation, we examine the relation between $\log({\rm L}_{\rm SB})$ and $\log({\rm L}_{acc})$.
\begin{figure}
\center
\includegraphics[width=0.5\textwidth]{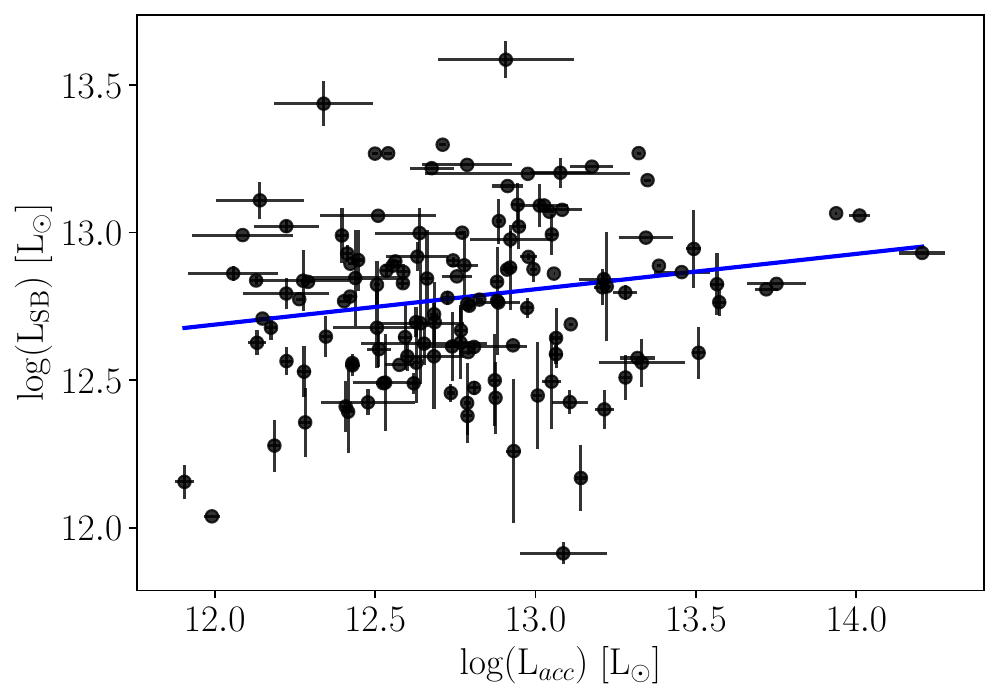}
\caption{L$_{\text{SB}}$ as a function of the L$_{acc}$. 
The blue line shows maximum-likelihood regression including intrinsic scatter. See the main text for details.} 
\label{fig:lsblacc}
\end{figure}

We modelled the relation using a maximum-likelihood linear regression that incorporates measurement uncertainties in both variables alongside an intrinsic scatter term, $\sigma_{int}$, treated as a free parameter. The relation was parametrised as
\begin{equation}
\log({\rm L_{SB}}) = a+b\log({{\rm L}_{acc}})
\end{equation}
\noindent
with the luminosities in units of L$_{\odot}$, where the intercept, $a$, slope, $b$, and intrinsic scatter, $\sigma_{int}$, were fitted simultaneously by maximising the joint likelihood function.
The resulting best-fitting relation is
\begin{equation}
\log(\mathrm{L}_{\mathrm{SB}}) = (11.25\pm0.71) + (0.12\pm0.06)
\log({\rm L}_{acc}/{\rm L}_{\odot})
\end{equation}
\noindent
with an intrinsic scatter of $\sigma_{int} = 0.264\pm0.017~\mathrm{dex}$.

The best-fitting slope ($b=0.12\pm0.06$) indicates only a weak dependence of L$_{\rm SB}$ on L$_{acc}$. To evaluate whether this shallow slope represents a statistically significant improvement over a null hypothesis of no correlation ($b=0$), we assessed the likelihood-ratio statistics 
$2 \left( \ln \mathcal{L}^*_{\rm{fit}} - \ln \mathcal{L}^*_{\rm{null}} \right),$
where $\mathcal{L}^*_{\rm{fit}}$ and $\mathcal{L}^*_{\rm{null}}$ denote the maximum likelihoods of the fitted model and the null model, respectively. We obtain $2\Delta\ln\mathcal{L}=4.6$, indicating that the model that includes a slope is only marginally preferred over a constant relation ($\sim2\sigma$ significance). The inferred dependence is shallow: an increase of one order of magnitude (1dex) in L$_{acc}$ corresponds to merely a factor of $\sim1.3$ increase in L$_{\rm SB}$. In contrast, the intrinsic scatter ($\sigma_{int}$=0.264) accounts for a factor of $\sim1.8$ dispersion in L$_{\rm SB}$ at fixed L$_{acc}$, demonstrating that the object-to-object variance dominates over any underlying scaling.

Overall, our results indicate that star formation and SMBH accretion are not tightly coupled on the timescales probed by these observations. Although a weak positive trend is marginally favoured statistically, the relation is dominated by intrinsic scatter, implying that FIR-bright quasars with similar accretion luminosities inhabit host galaxies spanning a wide range of starburst luminosities. We therefore find no evidence that the instantaneous accretion luminosity strongly regulates the recent star formation activity of the host galaxies. Furthermore, the estimated L$_{\rm SB}$ values, and consequently derived SFRs, are independent of the quasar environment as characterised by the presence of secondary counterparts within the SPIRE beam.

\subsection{UV/optical signatures}
\label{sec:optical}

\begin{figure*}
\center
\includegraphics[width=.48\textwidth]{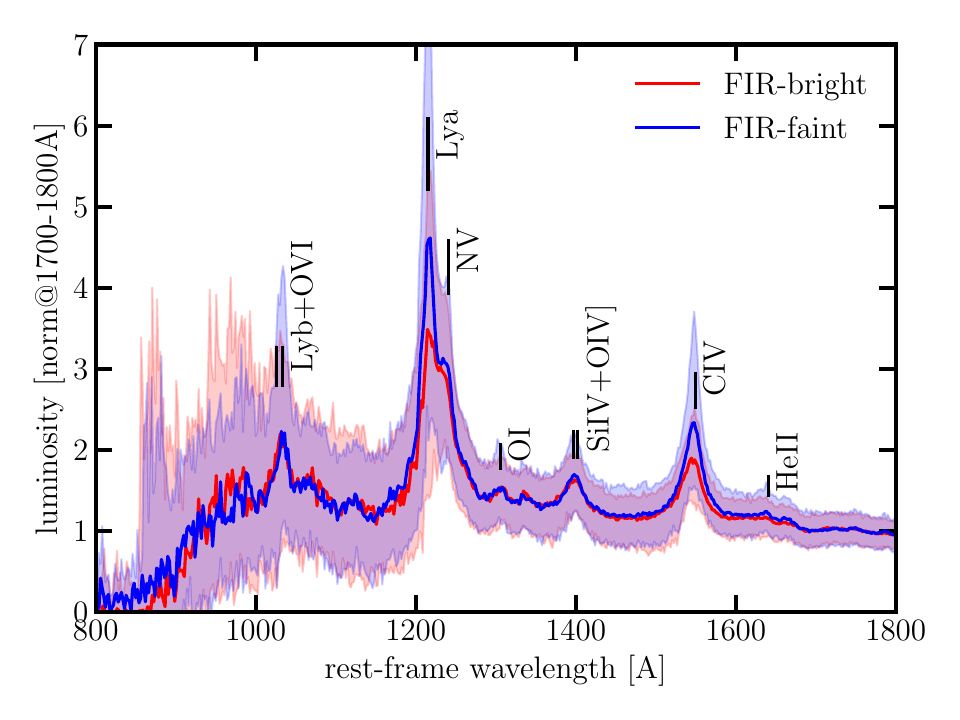}
\includegraphics[width=.48\textwidth]{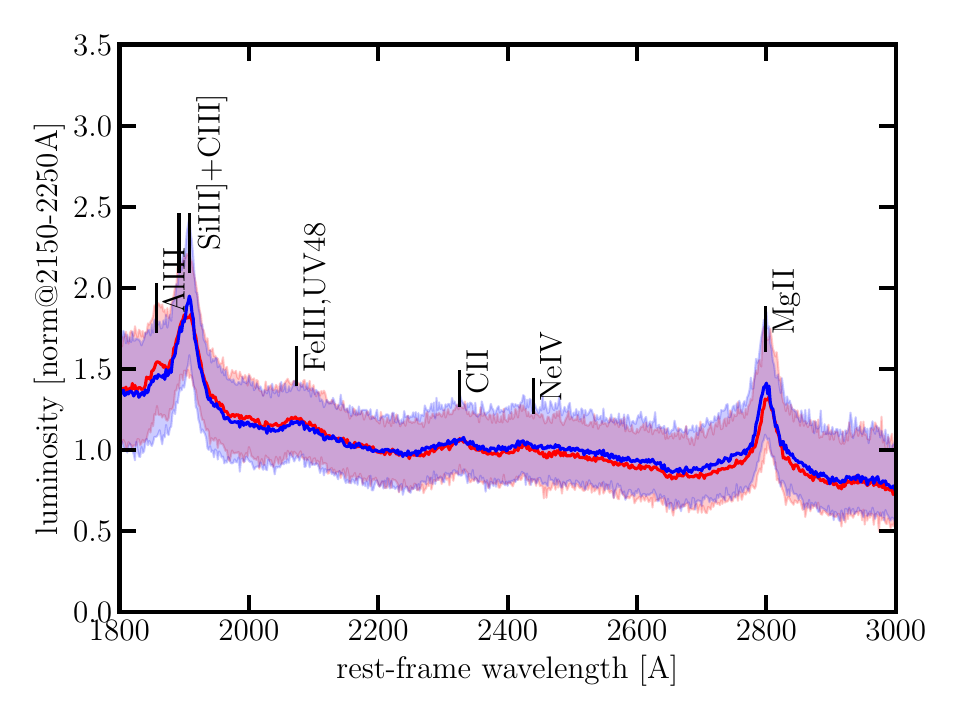}
\\
\includegraphics[width=.48\textwidth]{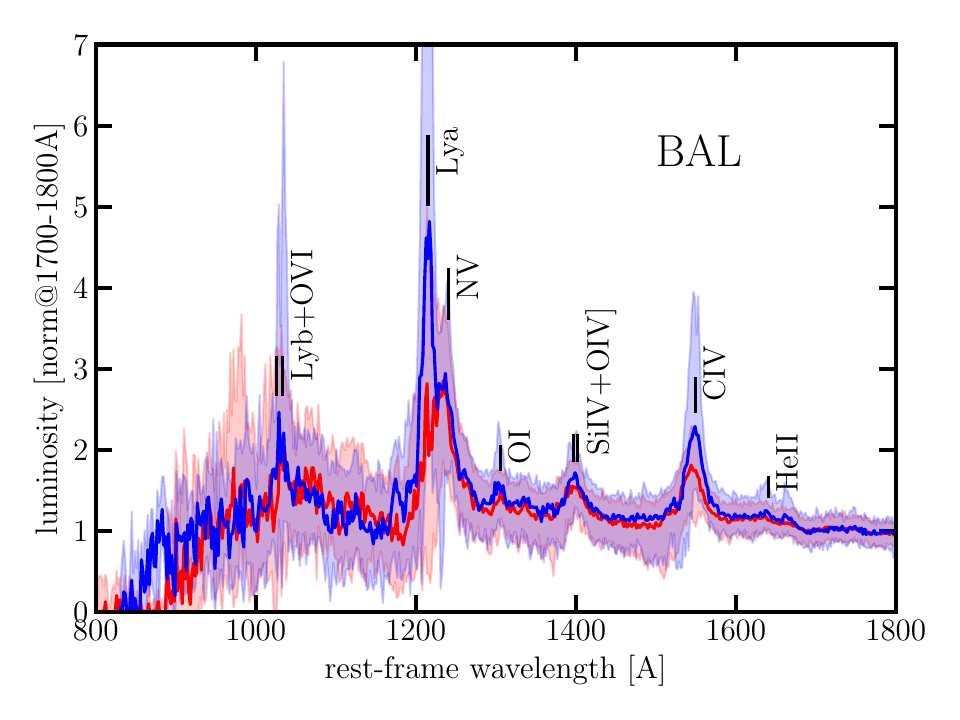}
\includegraphics[width=.48\textwidth]{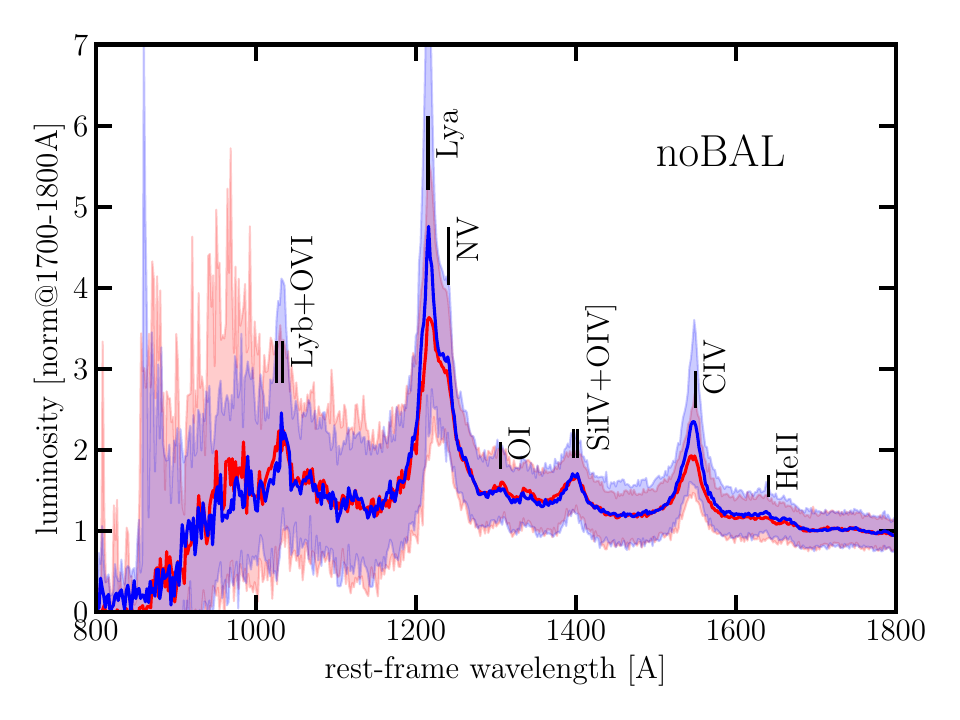}
\caption{Top row: Median stacked SDSS spectra from the FIR-bright quasars in our sample (red) and those from a matched in redshift and M$_{i}$ sample of FIR-faint (i.e. non detected individually by {\it Herschel}/SPIRE) quasars in the Stripe 82 field (blue). A minimum of 50, 107 and 101 quasars contribute to the \Lya, \CIV\ and \MgII\ features, respectively for either sample. The shaded region shows the range between the 16$^{th}$ and 84$^{th}$ percentiles. Bottom row: Blue part of the stacked spectra divided among BAL and non-BAL quasars.}
\label{fig:opticalspectra}
\end{figure*}

With the FIR and submm counterparts now unambiguously associated with the SDSS FIR-bright quasars in our sample, to gain insights into their nature we compare their UV/optical properties to those of FIR-faint quasars, i.e. quasars that are not individually detected by {\it Herschel}/SPIRE. We build a sample of 142 FIR-faint quasars in the HerS field, matched to the ALMA/SDSS quasar sample in redshift and $i$-band absolute, M$_{i}$. For the two samples, we constructed median stacked SDSS spectra (Fig. \ref{fig:opticalspectra}, top panel). The spectra were normalised at 1700--1800 \AA\ (2150--2250 \AA\ for the figure in the top right panel). 

We find indeed that \Lya\ and \CIV\ are stronger in the stacked spectrum of the FIR-faint quasars.  
At the same time, other high signal-to-noise features, such as \CIII\ and \MgII, as well as the continuum show little to no differences between the FIR-bright and FIR-faint quasar samples.

Splitting the two samples into non-BAL and BAL quasars (the latter corresponding roughly to about 25\% of each sample), we find that the contrast between FIR-bright and FIR-faint quasars persists, with the difference in the peak emission of \Lya\ being most pronounced in the BAL spectra (see Figure \ref{fig:opticalspectra}, bottom row). Note, however, separating the samples in BAL and no-BAL leaves the high-redshift bin underpopulated, resulting in a larger statistical uncertainty around the \Lya\ portion of the spectrum, rendering the associated results statistically less robust.

\begin{figure*}
\center
\includegraphics[width=.33\textwidth]{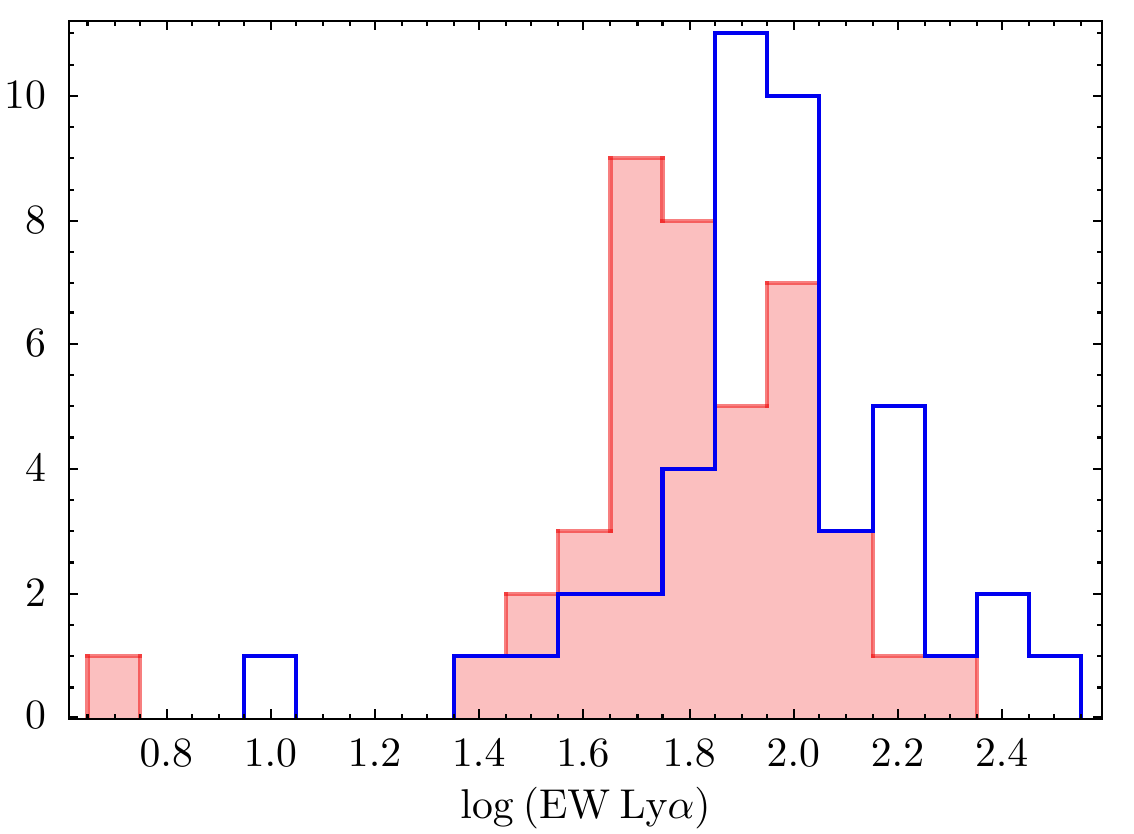}
\includegraphics[width=.33\textwidth]{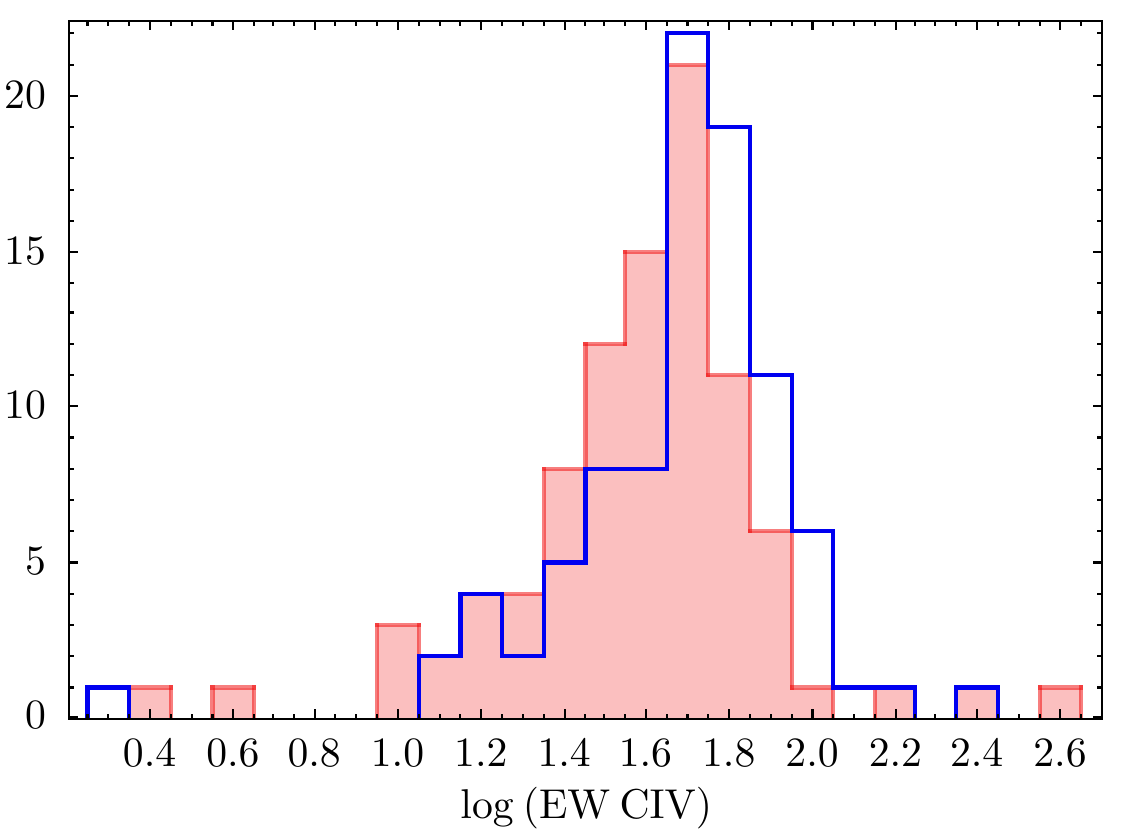}
\includegraphics[width=.33\textwidth]{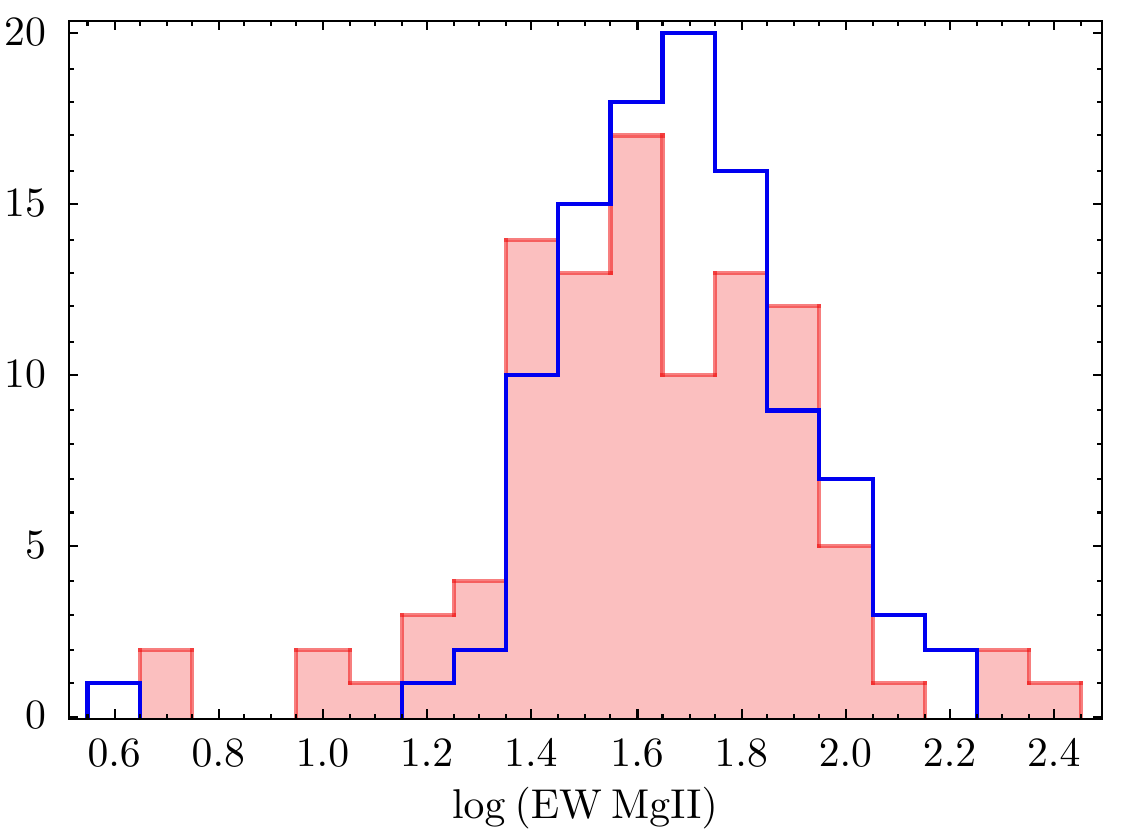}
\caption{The distribution of equivalent widths for \Lya, \CIV\ and \MgII\ for the FIR-bright (red) and FIR-faint (blue) quasar samples (EW values from \citealt{rakshit20}).}
\label{fig:ews}
\end{figure*}

\cite{li20} quantified variations of the \Lya\ equivalent width (EW) between submm-detected and undetected quasars with SCUBA2. They  confirmed earlier findings suggesting that quasars with submm detections tended to have weaker UV lines compared to quasars with no submm detections. They also reported that there was no indication that these quasars had redder optical continua compared to quasars with no submm counterparts. To explain the weaker lines, they invoked a scenario where a shielding gas blocks the ionisation continuum \citep{wu11} or an alternative case of a slow BLR development \citep{hryniewicz10}.

To build on this work, we use the emission line EWs from \cite{rakshit20}. A detailed analysis of the UV/optical spectral features is beyond the scope of this work. We therefore rely on pipeline measurements provided in the above work, with the corresponding uncertainties inherent to such measurements. 

Fig. \ref{fig:ews} shows the EW distributions for \Lya, \CIV\ and \MgII\ for the two samples (FIR-bright in red, FIR-faint in blue). We compared the distributions of the three emission lines across the two datasets using two-sample Kolmogorov–Smirnov (KS) tests.  \CIV\ (D = 0.245, p = 0.0067) and \Lya\ (D = 0.3390, p = 0.011) exhibited statistically significant differences. In contrast, the \MgII\ distributions are consistent with being drawn from the same parent population (D = 0.144, p = 0.21). As the FIR-bright and FIR-faint quasars were matched in M$_{i}$, to mitigate possible biases arising from emission-line contribution to the M$_{i}$, we repeated the KS test after excluding sources in the redshifts ranges  1.5 - 2 and 3.5 - 4.5 for the \MgII\ and \CIV\ EW distributions, respectively. The resulting p-values remain consistent with the original analysis, yielding 0.1967 for \MgII\ and 0.0026 for \CIV. 

To rule out the possibility that the observed behaviour is driven by the Baldwiin effect, i.e. the decrease of the EW with increasing continuum luminosity \citep{baldwin77} or with Eddington ratio \citep{wandel99}, we also performed two-sample KS tests on the bolometric luminosities, the rest-frame continuum luminosities at 1350 and 3000 \AA\ and the Eddington ratios (all quantities taken from \citealt{rakshit20}). The resulting p-values (0.5898, 0.8332, 0.8349 and 0.1775, respectively) are consistent with the two samples been drawn from the same parent population.

The difference in EW behaviour must therefore come from the physical conditions in the vicinity of the SMBH of the quasars. While the high-ionisation \CIV\ and resonant line \Lya\ are often produced in high-velocity, hot outflows, \MgII\ is a resonance line that traces colder, denser gas, often showing in-filled profiles compared to high-ionisation lines. The in-filling implies that the EWs remains relatively constant or only slightly decreases in the presence of outflows \citep{prochaska11}. \CIV\ on the other hand, is more sensitive to high-velocity wind dynamics \citep{coatman16}. Similarly, \Lya\ EWs are dictated by the distribution of neutral hydrogen and outflows \citep{verhamme15} and the less pronounced emission peak in the FIR-bright quasars might be indicative of residual neutral gas in the host.

\subsection{The nature of FIR-bright quasars}

Our multi-wavelength analysis of 142 FIR-bright, optically unobscured quasars ($1 \le z \le 4$) reveals a self-consistent picture of intense star formation coexisting with rapid SMBH growth. 

ALMA 870 \mums observations show that the submm continuum is largely unresolved or marginally resolved at 0.8\arcsec\ resolution, constraining these dusty starbursts to compact, nuclear regions ($\le6$ kpc). The compact nature of the 870 \mums emission is consistent with the in-situ assembly of the stellar bulge, a process seen at in higher resolution observations os submm populations spanning a comparable redshift range \citep{simpson15,tadaki17}. 

However, this interpretation does not exclude a contribution from merger-driven processes. Approximately 35\% of the sources in our sample have multiple ALMA counterparts within the SPIRE beam, which may indicate the presence of interacting systems or close companions. Establishing whether these components are physically associated with the quasar hosts requires redshift information for the individual ALMA sources, which is not available for the present sample. Therefore, while the compactness of the individual 870 \mums sources supports a scenario in which a significant fraction of the star formation occurs in compact nuclear regions, both in-situ growth and merger-related processes may contribute to the assembly of these systems, as postulated by e.g. \citealt{hopkins06} \citep[see also the review by][and references therein]{hodge20}.

The placement of these sources on the L$_{\rm SB}$ - L$_{\rm 1.4 GHz}$ plane demonstrates that, in their majority, they follow the standard FIR-radio correlation for star-forming galaxies. This suggests that the physical processes linking dust heating and cosmic ray production remain consistent in these systems, with FIR and radio energetics being governed by the starburst rather than the luminous AGN.
This interpretation is further reinforced by our CO line diagnostics (for details see Paper I). The detection of mid-$J$ transitions, specifically CO(6–5) and CO(7–6), contrasted with the absence of higher-$J$ lines, indicates an interstellar medium that is moderately excited by stellar radiation fields. The lack of highly excited components characteristic of AGN-driven X-ray dominated regions suggests that while the quasar dominates the bolometric luminosity, the host's massive molecular gas reservoirs are primarily heated by the ongoing compact starburst.

The BLR properties of the FIR-bright quasars diverge from those of their FIR-faint counterparts. While the \MgII\ and optical continuum remain comparable, the high-ionisation \CIV\ and resonant \Lya\ lines are notably suppressed in both flux and equivalent width. This behaviour is compatible with the presence of disk winds or outflows affecting the properties of the BLR, while \Lya\ might also be affected by the presence of lingering neutral hydrogen gas.

Taken together, the above findings suggest that FIR-bright quasars represent a short-lived transitional phase between the ``blowout'' stage of galaxy evolution and the emergence of unobscured, blue quasars \citep[e.g.][]{hopkins08}. The rarity of FIR-bright quasars, which comprise only a few per cent of the overall SDSS quasar population \citep{caoorjales12, pitchford16}, further supports the brevity of this evolutionary phase. In this picture, the systems are hosted by galaxies that remain exceptionally rich in gas and dust, providing a substantial reservoir of material that can fuel both rapid accretion onto the central SMBH and intense star formation that builds the host's bulge. This means that the quasar hosts are at the peak of their mass assembly while the SMBH has reached its peak growth rate. 
In this scenario, the now optically visible quasar has cleared the surrounding dusty envelope, while the circumnuclear molecular gas has yet to be fully depleted or expelled. The environmental diversity of our sample, which includes both isolated systems (65\%) and those with nearby companions (35\%), as reported in Paper I, further indicates that this extreme evolutionary phase is not tied to a single triggering mechanism. Instead, it can arise from both late-stage mergers and secular processes at work in already coalesced systems. 

\section{Summary}
\label{sec:sum}

In this work, we used the multi-wavelength properties of $1 \le z \le 4$ optically unobscured, FIR-bright quasars to bring together information regarding their star formation activity, accretion energetics and BLR conditions in order to build the evolutionary history of these rare population. Integrating high-resolution ALMA Band 7 photometry resolves {\it Herschel}/SPIRE source confusion, systematically lowering SED-fitting-derived SFRs by $\sim$20\%, while narrowing SFR uncertainties by $\sim$18\%. Despite these downward revision, FIR-bright quasars remain extreme starburst systems. Their FIR-to-radio properties and mid-$J$ CO excitation patterns confirm that star formation, rather than AGN heating, powers the dust emission and interstellar medium excitation, primarily within compact nuclear regions ($\le 6$ kpc). 

Comparing starburst and accretion luminosities reveals only a weak trend between L$_{\rm SB}$ and L$_{acc}$. Although a shallow positive correlation is marginally favoured statistically over a null hypothesis model, the intrinsic scatter dominates over the inferred trend, indicating that FIR-bright quasar hosts exhibit a wide range of starburst luminosities at a given accretion luminosity. These results suggest that SMBH accretion and star formation in the host are not coupled on the timescales probed by the observations.

A comparisons of their optical (SDSS) spectra with those of matched FIR-faint quasars show suppression in the EWs of high-ionisation \CIV\ and resonant \Lya, while low-ionisation \MgII\ lines and UV/optical continua remain largely unaffected. This signature points to high-velocity AGN winds and attenuation by lingering host neutral gas.

Synthesising these findings places 
FIR-brigh quasars in a brief, transitional phase between the heavily obscured ``blowout'' stage and unobscured quasars. These systems thus provide a critical laboratory for probing the host galaxy transformation, offering key insights into SMBH growth, gas depletion and feedback operating in tandem during one of the most transformative periods of galaxy evolution.
Featuring both isolated systems (65\%) and possibly interacting companions (up to 35\%), this rare population represents the simultaneous peak of host bulge assembly and SMBH accretion, triggered by diverse pathways prior to cold gas depletion. 

Ultimately, our results demonstrate that multi-wavelength broad-band photometry alone is not sufficient to capture the complex interplay between SMBH growth and extreme star formation in the hosts. Rather than a simple, feedback-regulated co-evolution, the weak coupling and dominant intrinsic scatter indicate that localised gas dynamics and diverse evolutionary pathways dictate host galaxy behaviour during this short-lived, FIR-bright phase. 

\begin{acknowledgements}
    EH  received funding from the European Union’s Horizon Europe research and innovation program under grant agreement No. 101188037 (AtLAST2). EH also acknowledges support from the JAO visitor program for a visit to Chile during which part of the work was carried out. The IAC is an accredited Severo Ochoa Centre of Excellence supported by grant CEX2025-001609-S, funded by MICIU/AEI/10.13039/501100011033. RS acknowledges the support of Universidade de Pernambuco (UPE), Federal Rural University of Pernambuco (UFRPE), and  CAPES PrInt (Coordination for the Improvement of Higher Education Personnel, Institutional Internationalization Program). This work makes use of TOPCAT \citep{taylor05}, {\sc ipython} \citep{PerezGranger07}, {\sc numpy} \citep{Walt11}, {\sc matplotlib} \citep{Hunter07}, {\sc scipy} \citep{Virtanen20}, {\sc astropy} \citep[a community-developed core {\sc python} package for Astronomy,][]{Astropy13}, APLpy \citep[an open-source plotting package for Python,][]{Robitaille12}. We used AI-based language tools to improve the clarity and grammar of the manuscript. No AI was used to generate scientific content, results, or conclusions. The authors are responsible for all content. 
\end{acknowledgements}

\bibliographystyle{aa}
\bibliography{biblio.bib}

\end{document}